\documentclass[]{elsarticle}
\usepackage{amsmath,amssymb,bm,color}
\usepackage{graphicx}
\usepackage{graphicx}
\usepackage{hyperref}
\usepackage{color}
\journal{Annals of Physics (published)}

\begin{document}

\title{Bose-Bose gases with nonuniversal corrections to the interactions:
a droplet phase} 
\author{Emerson Chiquillo}
\ead{emerson.chiquillo@unesp.br}
\address{Universidade Estadual Paulista (UNESP), Instituto de F\'isica
Te\'orica, S\~ao Paulo}

\begin{abstract}

Through an effective quantum field theory within Bogoliubov's framework and 
taking into account nonuniversal effects of the interatomic potential we 
analytically derive the leading Gaussian zero- and finite-temperature corrections
to the equation of state of ultracold interacting Bose-Bose gases. 
We calculate the ground-state energy per particle at zero and low temperature for
three- two- and one-dimensional two-component bosonic gases.
By tuning the nonuniversal contribution to the interactions we address and 
establish conditions under which the formation and stability of a self-bound 
liquidlike phase or droplet with nonuniversal corrections to the interactions
(DNUC) is favorable.
At zero temperature in three-dimensions and considering the nonuniversal
corrections to the attractive interactions as a fitting parameter the energy per
particle for DNUC is in good agreement with some diffusion Monte Carlo results.
In two dimensions the DNUC present small deviations regarding conventional 
droplets. 
For the one-dimensional DNUC the handling of the nonuniversal effects to the 
interactions achieves a qualitative agreement with the trend of some available
Monte Carlo data in usual droplets.
We also introduce some improved Gross-Pitaevskii equations to describe 
self-trapped DNUC in three, two and one dimension.
We briefly discuss some aspects at low temperature regarding nonuniversal
corrections to the interactions in Bose-Bose gases. 
We derive the dependencies on the nonuniversal contribution to the interactions
but also on the difference between intra- and inter-species coupling constants.
This last dependence crucially affect the three- and the two-dimensional DNUC 
driving thus to a thermal-induced instability.
This thermal instability is also present in one-dimensional Bose-Bose gases, but 
it is not relevant on the formation of DNUC.
This is explained because the necessary attraction mechanism to achieve this
phase naturally arises in the fluctuations at zero temperature without major
restrictions as it happens in the other dimensions.

\end{abstract}

\maketitle

\section{Introduction}

The highly nontrivial and subtle stabilization between the mean-field energy and 
the energy coming from the fluctuations in ultradilute Bose-Bose gases has 
attracted an increasing interest.
Such a balance has given rise to the liquidlike phase or droplet phase, revealing
thus the crucial role played by the fluctuations \cite{Petrov1,Petrov2}.
This intriguing new  state of matter lies in the mechanical stabilization of 
paradigmatic mean-field collapse threshold in a condensed Bose-Bose mixture 
through the first Gaussian correction coming from zero-point motion of Bogoliubov
excitations.
This proposal has allowed the birth of a fascinating and fast development
research field
\cite{1D-3D-mixtures, Drop-1d,Swirling,Bose-Bose,Rabi-E,Review,Adhikari2,
Collective-excit,Pairing1,Pairing2,Phonon1,Normal-superfluid,Boudjem,Bose-Bose2,
Evaporation,Phonon2,Spinor1,Spinor2, Criticality}. Though the fluctuations are 
usually small and these are generally neglected, recent experimental achievements
have put in evidence the outstanding influence of these small corrections on the 
equation of state for attractive Bose-Bose gases.
The droplet phase has been obtained in homonuculear $^{39}$K-$^{39}$K mixtures
\cite{Exper-droplet1,Exper-droplet2,Exper-droplet3,Exper-droplet4} and 
heteronuclear $^{41}$K-$^{87}$Rb mixtures \cite{Exper-droplet5}.
Also has been studied the dynamical formation of self-bound droplets in an
attractive mixture of $^{39}$K atoms \cite{Exper-droplet6}.
More recently the observation of a Lee-Huang-Yang (LHY) fluid in a $^{39}$K spin 
mixture confined in a spherical trap potential has been achieved
\cite{Exper-droplet7}.
Droplets have been also obtained in a mixture of $^{23}$Na and $^{87}$Rb with 
tunable attractive inter-species interactions \cite{Exper-droplet8}.
It is also worth mentioning that soon after the theoretical proposal of
Bose-Bose droplets, this idea was extended to the context of one-component Bose
gases with magnetic atoms
\cite{trapped-dip-LHY-1,trapped-dip-LHY-2,d-LHY-1,d-LHY-2,Montecarlo-dip1,
d-LHY-3}.
Experimental evidence of dipolar droplets existence with atoms of $^{164}$Dy 
\cite{Rosensweig,Self-Drop-Dy,Trap-Drop-Dy-1,Trap-Drop-Dy-2} and $^{168}$Er
\cite{Erbium} was also discovered.
However, it is remarkable that comparing experimental evidence with theoretical 
predictions, in both droplets in mixtures of Bose-Bose gases and droplets in 
bosonic dipolar gases, some discrepancies have been found.
In fact, in three-dimensional Bose-Bose gases it was experimentally observed that
a stable droplet formation requires a critical minimum number of atoms
\cite{Exper-droplet1}. This critical value is less than predicted by the 
proposed model in Ref. \cite{Petrov1}.
Some differences are also present comparing the theoretical model with numerical 
results obtained by means of Monte Carlo methods
\cite{Numeric-drop,Montecarlo-Bose-Bose1,Montecarlo-Bose-Bose2}.
Therefore, given the discrepancy between the theoretical model and both 
experimental and numerical results, it is necessary to find a model that allows a
better description of this phenomenon.
In an attempt to achieve this, a phenomenological and beyond LHY framework has
been developed in Ref. \cite{Beyond LHY} for three- and one-dimensional 
Bose-Bose droplets. 

From a general theoretical framework, such a droplet phase in both two-component 
ultracold Bose gases and dipolar bosonic gases remains weakly interacting,
allowing thus for a theoretical perturbative treatment.
A key ingredient in Bose-Bose droplets description is to consider the weakly
interatomic interactions approximated by a local contact interaction. In this
sense, only one parameter characterizes the two-body interactions and it is
encoded in an effective theory in terms of the s-wave scattering length $a$.
This unique dependence is considered as an universal regime \cite{Andersen}. On 
the other hand, the fact that physical quantities depend on properties other than
the s-wave scattering length is considered as a nonuniversal regime.
In addition and due to the sensitive nature of the fluctuations themselves the
effect on these of some other parameters remains as an open issue.
Therefore for a better understanding of fluctuations behavior on interacting 
Bose-Bose mixtures and motivated by the enhanced role of these, we consider an
extra ingredient.
We propose a step beyond by researching the effect of the nonuniversal 
corrections to the interatomic interaction potential  \cite{Nonuniversal1}.
We address and focused theoretically on formation and stability of self-bound
Bose-Bose droplets with nonuniversal corrections to the interactions (DNUC) both
at zero and low temperature. 
We consider DNUC in three- two- and one-dimension.
Thermal corrections on mixtures of two-component bosonic gases with zero-range
interactions have been studied in Ref. \cite{1D-3D-mixtures} (see also
\cite{Bose-Bose}).
Recently, the implementation of Monte Carlo algorithms has been carried out in 
the study of finite temperature in Bose-Bose mixtures
\cite{Montecarlo-finite-T-1,Montecarlo-finite-T-2,Montecarlo-finite-T-3}.
We stress that although in recent years have been included nonuniversal 
corrections to the interactions on the study of the equation of state in 
one-component Bose gases in three \cite{Non-univ-3d}, two \cite{Non-univ-2d} and 
one dimension \cite{Non-univ-1d}, the necessary conditions for a droplet 
formation have not been considered.

The rest of the paper is organized as follows.
In Section \ref{Model and preliminaries} we introduce an effective-field theory 
for interacting bosons taking into account nonuniversal corrections to the 
interactions.
We also bring up some elements of interest in the path-integral formalism which 
are employed to obtain the equation of state of two-component bosonic gases with
nonuniversal corrections to the interactions in dimension $d$, with $d=3,2,1$.
So, we derive the mean-field grand-canonical potential. We also obtain general
expressions for the grand-canonical potential dealing with fluctuations and 
thermal effects.
In order to simplify the problem and to know some insights into the underlying
physics of the DNUC we consider a symmetric scenario.
To keep the notation as simple as possible, in all dimensions we use $a$ for the
scattering length and $r$ for the effect of the nonuniversal contribution to the
interactions, respectively.
After establishing some initial aspects, we calculate and analyze the existence
conditions of stable $3d$ DNUC at zero and finite temperature in section
\ref{Three-dimensional model}.
We present two different equations of state each related to two different
expansions of the phase shift in $3d$ scattering theory.
In section \ref{Two-dimensional model} we investigate the stable formation of
$2d$ DNUC.
We also include finite-temperature corrections.
In section \ref{One-dimensional model} we present the formation of stable $1d$ 
DNUC at zero and finite temperature.
In this section we also discuss some interesting differences regarding $2d$ and 
$3d$ DNUC.
In section \ref{Summary and outlook} we draw a final discussion and we present 
some future perspectives of our results.

\section{Model and preliminaries}
\label{Model and preliminaries}
\subsection{An effective-field theory for interacting bosons}

At theoretical level, in the regime of low-momentum expansion of the interaction
potential some improvements to obtain useful and analytical expressions
describing ultracold bosons have been proposed. 
In a first attempt within the framework of an effective field theory, the 
next-to-leading effects of the interactions in homogeneous three-dimensional 
Bose-Einstein condensates are calculated in Ref. \cite{Nonuniversal1}.
An effective Lagrangian constrained by the symmetries of the original Lagrangian, 
namely, Galilean invariance, parity, and time reversal, is considered. 
The coefficients of this Lagrangian are determined by imposing that the effective
theory reproduces the amplitude for low-energy atom-atom scattering up to errors
that scale like the square of the energy.
Going to the center of momentum frame, using the energy conservation and 
regularization prescriptions, the exact T-matrix element can be determined 
analytically.
Thus the scattering length and the nonuniversal corrections for the s-wave 
scattering are defined by the expansion of the phase shift.
On the other hand, in Refs. \cite{Nonuniversal2,Non1,Non2} the next-to-leading 
order term in the low-momentum expansion of the interaction potential is obtained
based on the calculation of the energy shift due to a phase shift in the 
wavefunction.
We stress that due to the different considerations employed, the two above 
approaches do not coincide with each other.
In particular, in the last approximation if the nonuniversal correction goes to
zero some finite contributions of the scattering length still, and we not obtain 
the conventional dependence on the scattering length.
It is worth noting that the improvement developed in Ref. \cite{Nonuniversal1}
has been employed in recent years to study the effects of the nonuniversal
contribution to the interactions in the equation of state of bosonic gases in
three dimensions \cite{Non-univ-3d}, with results in good agreement with Monte 
Carlo calculations. 
Such a effective field theory has been also adopted in two \cite{Non-univ-2d} and
one dimension \cite{Non-univ-1d}.
So we consider the effective field theory approach from Ref. \cite{Nonuniversal1}.
Our motivation lies in the fact that this procedure is directly related to the
scattering theory by means of the T-matrix. Furthermore when the nonuniversal 
effects vanishes we recover the widely known results in the study of bosons and 
employed in Bose-Einstein condensation research. 
Additionally we can also extent the idea to lower dimensions.
So we consider a formalism reasonably suitable to study an effective field theory
for ultracold bosons.
In order to establish the effective Lagrangian for two component bosonic gases 
with nonuniversal corrections to the interactions, for simplicity, we first 
consider one-component Bose gases. After that, we adopt the natural extension to 
two-component bosonic gases.

In the path-integral formalism we consider interacting and identical bosons of 
mass $m$ with chemical potential $\mu$ in a $d$-dimensional, $(d=3,2,1)$, box of 
volume $L^d$ with the Lagrangian density given by
\begin{eqnarray}
\mathcal{L} &=&
\psi^* (\textbf{r},\tau) \big(\hbar\frac{\partial}{\partial\tau}
- \frac{\hbar^2}{2m}\nabla^2  -\mu\big)\psi(\textbf{r},\tau),
\nonumber
\\
&+&\frac{1}{2}\int d^d r'|\psi(\textbf{r}',\tau)|^2
V(|\textbf{r}-\textbf{r}'|) |\psi(\textbf{r},\tau)|^2.
\label{L1}
\end{eqnarray}
The fields are considered at position $\mathbf{r}$ and imaginary time $\tau$, and
$V(|\textbf{r}-\textbf{r}'|)$ is the two-body interaction potential between 
bosons.
Dealing with ultracold and dilute bosons, the most common scheme to taking into
account interaction effects is replacing $V(r)$ 
(where $r\equiv|\textbf{r}-\textbf{r}'|$),
adopting an approximated zero-range Fermi pseudo-potential $V^{(0)}(r)=g^{(0)}
\delta^{(d)}(r)$. 
The specific form of the strength of the interaction $g^{(0)}$ depends on the 
dimension of the system.
At the moment the most relevant aspect of this term is that it only depends on 
the scattering length $a$.
An improvement of the zero-range approximation can be achieved by replacing the
Fourier transform of the interaction potential 
$\widetilde{V}(k)=\int d^dr\exp{(\text{i}\textbf{k}\cdot\textbf{r})} V(r) $
with the pseudo-potential $\widetilde{V}(k)= g^{(0)} + g^{(2)}k^2$.
This is obtained as the Taylor expansion in $k$ around $k=0$ of 
$\widetilde{V}(k)$ up order two. 
So we have $g^{(0)}=\widetilde{V}(0)$, and 
$g^{(2)}=(1/2)\widetilde{V}''(k)|_{k=0}$.
Since $k=0$ is a minimum of the potential, the linear contribution in the Taylor
expansion vanishes. By means of the inverse Fourier transform of the 
pseudo-potential $\widetilde{V}(k)$, the pseudo-potential in real space is used
into the second line of Eq.~(\ref{L1}), and we get the following effective 
Lagrangian density including effective nonuniversal effects of the interactions
\begin{eqnarray}
\mathcal{L} &=&
\psi^* (\textbf{r},\tau) \big(\hbar\frac{\partial}{\partial\tau}
- \frac{\hbar^2}{2m}\nabla^2  -\mu\big)\psi(\textbf{r},\tau),
\nonumber
\\
&+&\frac{1}{2} g^{(0)}|\psi(\textbf{r},\tau)|^4
- \frac{1}{2} g^{(2)}|\psi(\textbf{r},\tau)|^2\nabla|\psi(\textbf{r},\tau)|^2.
\label{L1-1}
\end{eqnarray}
Like $g^{(0)}$, the strength of the nonuniversal effects to the interactions 
$g^{(2)}$ also depends on the spatial dimension of the system and the specific 
form will be discussed later.

By considering the natural extension of the procedure outlined above we consider 
two-component interacting ultracold Bose gases. 
Each component is described by a complex bosonic field $\psi_\alpha$ 
($\alpha=1,2$) with equal-mass $m$, and chemical potentials $\mu_{\alpha}$.
The effective Euclidean Lagrangian density in a $d$-dimensional box of volume
$L^d$ is given by
\begin{eqnarray}
\mathcal{L}=\mathcal{L}_0 + \mathcal{L}^{(0)}_{\text{int}}
+\mathcal{L}^{(2)}_{\text{int}},
\label{L}
\end{eqnarray}
where
 \begin{eqnarray}
\mathcal{L}_0 = \sum_{\alpha=1,2}
\Big[\psi_\alpha^* (\textbf{r},\tau) \big(\hbar\frac{\partial}{\partial\tau}
- \frac{\hbar^2}{2m}\nabla^2  -\mu_\alpha\big)\psi_\alpha(\textbf{r},\tau)\Big],
\end{eqnarray}
\begin{eqnarray}
\mathcal{L}^{(0)}_{\text{int}} 
= \frac{1}{2}\sum_{\sigma=1,2}g^{(0)}_{\alpha\sigma}
|\psi_\alpha(\textbf{r},\tau)|^2|\psi_\sigma(\textbf{r},\tau)|^2,
\end{eqnarray}
and,
\begin{eqnarray}
\mathcal{L}^{(2)}_{\text{int}} =
- \frac{1}{2}\sum_{\sigma=1,2}g^{(2)}_{\alpha\sigma}
|\psi_\alpha(\textbf{r},\tau)|^2 \nabla^2 |\psi_\sigma(\textbf{r},\tau)|^2.
\label{Lagrangian-FR}
\end{eqnarray}
The specific form of the couplings will defer further discussion until
addressing each dimension separately.
The strength of the zero-range interactions is expressed through the intra- 
and interspecies coupling constants $g^{(0)}_{\alpha\alpha}$ and
$g^{(0)}_{\alpha\sigma}$, respectively.
The role played by the effects beyond the zero-range approximation to the 
interactions is encoded in expression (\ref{Lagrangian-FR}).
Thus we have the coupling constants for intra-species and inter-species 
considering the nonuniversal corrections to the interactions 
$g^{(2)}_{\alpha\alpha}$ and $g^{(2)}_{\alpha\sigma}$, respectively. 

In order to obtain the ground state of the system, we calculate the grand 
potential $\Omega = -k_B T \log\mathcal{Z}$. 
In the path-integral formalism the grand canonical partition function
$\mathcal{Z}$ of the system at temperature $T$ is written as
$\mathcal{Z} = \int \mathcal{D}[\Psi,\Psi^*]\text{exp}(-S[\Psi,\Psi^*]/\hbar),$
and it is governed by the action
\begin{eqnarray}
S[\Psi,\Psi^*]=\int_0^{\hbar\beta} d\tau \int_{L^d}  d^d r
\mathcal{L} [\Psi,\Psi^*],
\end{eqnarray}
with $\Psi=(\psi_1,\psi_2)^T$, $\beta = (k_B T)^{-1}$, $k_B$ the Boltzmann 
constant, $d=3,2,1$, and $\mathcal{L}$ given by Eq. (\ref{L}).

At zero temperature we assume the bosons condensate into the zero-momentum 
states. Although strictly Bose-Einstein condensation is prevented in low
dimensionality, a finite-size system at sufficiently low temperature allows for a
quasicondensation \cite{quasi1,quasi2}.
In order to perform a perturbative expansion we set
$\psi_\alpha (\textbf{r},\tau) = \phi_{\alpha}(\textbf{r})
+ \eta_\alpha (\textbf{r},\tau)$. Where
$\phi_{\alpha}(\textbf{r}) 
\equiv \langle \psi_\alpha (\textbf{r},\tau) \rangle$ 
\cite{Stoof},
corresponds to the macroscopic condensate ($3d$) or quasicondensate ($2d$, and
$1d$) wave-function.
The fluctuations around $\psi_{\alpha}$ are given by
$\eta_\alpha (\textbf{r},\tau)$ and these are orthogonal to the condensate
(quasicondensate) of the same species \cite{Stoof, Modes}.
Since here the relative phases of the wave-functions do not  play a role, we 
have $n_{\alpha}=|\phi_{\alpha}(\textbf{r})|^2=N_\alpha/L^d$, 
as the density of particles or condensate (quasicondensate) density in the 
mean-field approximation.
Since $\phi_{\alpha}(\textbf{r})$ describes the condensate (quasicondensate), the
linear terms in the fluctuations vanish such that $\phi_{\alpha}(\textbf{r})$ 
really minimizes the action \cite{Stoof}.
Therefore results in the mean-field approximation could be obtained by minimizing
the mean-field grand potential $\Omega_0$ with respect to $\phi$, as will be seen
later.
Now, by expanding the action up to the second order (Gaussian) in
$\eta_\alpha (\textbf{r},\tau)$ and $\eta^*_\alpha (\textbf{r},\tau)$, we arrive
at the split grand potential $\Omega= \Omega_0 + \Omega_{\text{f}}$. Where
$\Omega_0$ is the mean-field grand potential, while $\Omega_{\text{f}}$ takes 
into account both zero- and finite-temperature fluctuations to the 
grand-potential.

\subsection{Mean-field approximation to the grand-potential}

At first instance we carry out the analysis of the mean-field contribution to the
grand-potential, from which we get the following
\begin{eqnarray}
\frac{\Omega_0}{L^d}
= \sum_{\alpha=1,2} \big(-\mu_\alpha \phi^2_\alpha
+ \frac{1}{2}\sum_{\sigma=1,2}g^{(0)}_{\alpha\sigma}
\phi^2_\alpha \phi^2_\sigma\big),
\end{eqnarray}
which is not dependent on the nonuniversal effects to the interaction, given the
derivative term of $\mathcal{L}^{(2)}_{\text{int}}$.
So by minimizing $\Omega_0$ with respect to $\phi_{\alpha}$, such that
$ \partial\Omega_0 / \partial\phi_{\alpha}=0$, we obtain
\cite{1D-3D-mixtures},
$ \mu_\alpha = g^{(0)}_{\alpha\alpha}\phi^2_\alpha 
+ g^{(0)}_{\alpha\sigma}\phi^2_\sigma$, with
$\alpha,\sigma=1,2$ and $\alpha\neq\sigma$.
Hence the mean-field grand potential becomes
\begin{eqnarray}
\frac{\Omega_0}{L^d}
= -\frac{1}{2}\sum_{\alpha}
\frac{g^{(0)}_{\sigma\sigma}\mu_{\alpha}^2
- g^{(0)}_{\alpha\sigma}\mu_{\alpha}\mu_{\sigma}}
{g^{(0)}_{\alpha\alpha}g^{(0)}_{\sigma\sigma} -[g_{\alpha\sigma}^{(0)}]^2}
\label{Grandp0}
\end{eqnarray}
    
\subsection{Gaussian fluctuations to the grand-potential}

The grand potential of the Gaussian fluctuations $\Omega_{\text{f}}$ is provided 
by \cite{Regularization}
\begin{eqnarray}
\Omega_{\text{f}} = \frac{1}{2\beta}\sum_{k>0}
\sum^{+\infty}_{\substack{n=-\infty}} \ln \det [\mathbb{G}^{-1}(k,\omega_n)],
\label{Grand1}
\end{eqnarray}
with the bosonic Matsubara's frequencies $\omega_n=2\pi n/ (\hbar \beta)$, and
the $4\times4$ inverse fluctuation propagator $\mathbb{G}^{-1}$ given as
\begin{equation}
\mathbb{G}^{-1} =
\begin{pmatrix}
\textbf{G}_{11}^{-1} & \textbf{G}_{12}^{-1} \\
\textbf{G}_{12}^{-1} & \textbf{G}_{22}^{-1} \\
\end{pmatrix},
\label{inverse of gaussian propagator}
\end{equation}
with the symmetric $2\times2$ matrices
\begin{equation}
\textbf{G}_{11}^{-1} =
\begin{pmatrix}
-\text{i}\hbar \omega_n +\epsilon_k + h_{11}
& (g^{(0)}_{11} + g_{11}^{(2)}k^2)\phi^2_1 \\
(g^{(0)}_{11} + g_{11}^{(2)}k^2)\phi^2_1 &
\text{i}\hbar \omega_n +\epsilon_k + h_{11}\\
\end{pmatrix}
,
\end{equation}
$h_{11}=(2g_{11}^{(0)}+ g_{11}^{(2)}k^2)\phi^2_1
+ g_{12}^{(0)}\phi^2_2 - \mu_1$,
$\textbf{G}_{22}^{-1} = \textbf{G}_{11}^{-1}(1 \leftrightarrow2),$
and
\begin{equation}
\textbf{G}_{12}^{-1} =(g_{12}^{(0)} + g_{12}^{(2)} k^2) \phi_1\phi_2
\begin{pmatrix}
 1 & 1 \\
 1 & 1 \\
\end{pmatrix}
.
\end{equation}
By solving the determinant of the inverse propagator we get
\begin{eqnarray}
\Omega_{\text{f}}= - \frac{1}{2\beta}\sum^{+\infty}_{\substack{k>0\\
n=-\infty}} \ln \big[(\hbar^2\omega_n ^2 + E_+^2)
(\hbar^2\omega_n ^2 + E^2_-)\big],
\label{Grand1}
\end{eqnarray}
with the zero-point energy of collective excitations or Bogoliubov spectra given 
as
\begin{eqnarray}
E_\pm (k,\mu_{1},\mu_{2}) = [{\epsilon^2(k)}+
{\epsilon(k)f(\mu_{1},\mu_{2})}]^{1/2},
\end{eqnarray}
with the free-particle energy $\epsilon(k)=\hbar^2k^2/2m$, and
\begin{eqnarray}
f(\mu_{1},\mu_{2}) = \frac{1}{\Delta_-} \big(\tau_+ \pm \sqrt{\tau_-^2
+ 4\Lambda}\big),
\end{eqnarray}
where
$ \tau_+ = f^-_2 \mu_1 + f^-_1 \mu_2, \tau_- = f^+_2 \mu_1 - f^+_1 \mu_2,$
$\Lambda = h^2(\Delta_+\mu_1\mu_2 - h_2\mu^2_1 - h_1\mu_2^2),$
$ \Delta_{\pm} = g^{(0)}_{11}g^{(0)}_{22}  {\pm} [g^{(0)}_{12}]^2,$
$h=g^{(0)}_{12} + g^{(2)}_{12}k^2,\,
h_1=g^{(0)}_{11}g^{(0)}_{12},\,
h_2=g^{(0)}_{22}g^{(0)}_{12},$
$f^\pm_1 = g^{(0)}_{11}(g^{(0)}_{22} + g^{(2)}_{22}k^2)
\pm g^{(0)}_{12}(g^{(0)}_{11}+g^{(2)}_{11}k^2),$
$f^\pm_2=f^\pm_1(1\leftrightarrow 2).$
The sum over the bosonic Matsubara's frequencies given by Eq. (\ref{Grand1}) can
be read as
$\Omega_{\text{f}}= \Omega_g + \Omega^{(T)}_g$
\cite{Regularization, Le Bellac},
such that
\begin{eqnarray}
\Omega_g =\frac{1}{2}\sum_{k,\pm}E_{\pm}(k),
\end{eqnarray}
represents the Gaussian fluctuations at zero temperature, and,
\begin{eqnarray}
\Omega^{(T)}_g = \frac{1}{\beta} \sum_{k,\pm}
\ln \big(1-e^{-\beta E_\pm(k)}\big),
\label{Grand-T}
\end{eqnarray}
considers the thermal Gaussian fluctuations.
For now we will leave this last contribution aside and we focus on results at
zero temperature.
However, we will return to thermal effects up later.
In the continuum limit  $\sum_k\rightarrow L^d \int d^dk/(2\pi)^d$, the
contribution of the zero-temperature Gaussian fluctuations to the ground-state 
can be read as
\begin{eqnarray}
\frac{\Omega_g}{L^d} = \frac{S_d}{2}
\sum_{\pm}
\int_0^\infty \frac{dk}{(2\pi)^d}\, k^{d-1} E_\pm,
\label{grand-pot}
\end{eqnarray}
with $S_d=2\pi^{d/2}/\Gamma(d/2)$ the solid angle in $d$ dimensions and
$\Gamma(x)$ the Euler gamma function.
It is worth mentioning that although this integral is ultraviolet divergent at
$d=3,2,1$, it is possible to obtain finite results avoiding this issue by means
of regularization methods \cite{Regularization,Hooft,dim-reg}.

\subsection{Equal coupling constants considerations}

In order to obtain closed analytical expressions useful simplifications take 
place when the intra-species and the nonuniversal coupling constants are equal,
such that,
$g^{(0)}_{11}=g^{(0)}_{22}=g^{(0)}$ and, $g^{(2)}_{11}=g^{(2)}_{22}=g^{(2)}$. 
In addition we take the same density in different species $n_1=n_2=n/2$. 
Given these considerations, it is natural to take the same chemical potential,
$\mu_1=\mu_2=\mu$.
From these simplifications we get the following mean-field grand potential
\begin{eqnarray}
\frac{\Omega_0}{L^d} = -\frac{\mu^2}{g^{(0)} + g^{(0)}_{12}},
\label{grand-symmetric}
\end{eqnarray}
and the respective Bogoliubov spectra
\begin{eqnarray}
E^2_+
= (1+\delta_+ \mu)\Big(\frac{2\mu}{1+\delta_+ \mu} + \epsilon_k\Big)\epsilon_k,
\label{Bogol}
\end{eqnarray}
\begin{eqnarray}
E^2_-
= (1+\delta_- \mu)\Big(\frac{2\lambda\mu}{1+\delta_- \mu}
+ \epsilon_k\Big)\epsilon_k,
\end{eqnarray}
with
\begin{eqnarray}
\delta_{\pm} = \frac{4m}{\hbar^2}
\Big(\frac{g^{(2)}\pm g^{(2)}_{12}}{g^{(0)}+g^{(0)}_{12}} \Big),
\hspace{2cm}
\lambda = \Big(\frac{g^{(0)} - g^{(0)}_{12}}{g^{(0)}+g^{(0)}_{12}} \Big).
\label{constants}
\end{eqnarray}

\section{Three-dimensional model}
\label{Three-dimensional model}

It is well known from scattering theory that in three spatial dimensions the
s-wave scattering length $a$ and the s-wave effective range of interaction $r$
are the coefficients of the following expansion of the phase shift
$\delta(k)$ \cite{Scattering1,Scattering2}
\begin{eqnarray}
k\cot[\delta (k)] = - \frac{1}{a} + \frac{1}{2}rk^2 + \textit{O}(k^4)+...,
\label{fr}
\end{eqnarray}
where $k^2$ is the energy in units of $\hbar^2/2m$, and $m$ is the reduced mass.
In obtaining the couplings constants in Eq. (\ref{L}), we briefly describe some 
general aspects related to their calculation in a recently perspective 
\cite{On-shell}.
By adopting an on-shell approximation for the T-matrix equation, analytical
expressions connecting the Fourier transform of the interaction potential to 
the s-wave phase shift are obtained. In particular, explicit forms of the
low-momentum parameters in terms of the s-wave scattering length and the
effective-range of the interactions are calculated.
So the matrix element of the transition operator of scattering theory is
decomposed
into partial waves. The orthogonality of Legendre functions, and the uniqueness
of the representation in partial waves are used. Likewise,
it is chosen the term with  $l = 0$ (with $l$ the degree of Legendre polynomials)
i.e. it is selected the s-wave term. Finally, the on-shell approximation is
performed \cite{Scattering2}.
As a consequence, the matrix element or s-wave transition element takes the form
\begin{eqnarray}
T_{l=0}(k) = V_{l=0}(k) + V_{l=0}(k) C(k)  T_{l=0}(k),
\label{T-general}
\end{eqnarray}
with the Fourier transform of the spherically-symmetric interaction potential
$V(k)$ and, the dimensional dependent constant $C(k)$. In $3d$ case we have
\begin{eqnarray}
C(k) = -ik\frac{m}{4\pi\hbar^2}.
\label{C}
\end{eqnarray}
Moreover, the scattering theory shows that $T_{l=0}(k)$ is written in terms of
the s-wave scattering amplitude $f_{l=0}(k)$ as follows
\begin{eqnarray}
T_{l=0}(k) = -\frac{4\pi\hbar^2}{m} f_{l=0}(k).
\label{Tk}
\end{eqnarray}
In turn the s-wave scattering amplitude $f_{l=0}(k)$ is related to the s-wave
phase shift $\delta(k)$ by means of
\begin{eqnarray}
f_{l=0}(k) = \frac{1}{k[\cot[\delta(k)]-\text{i}]}.
\label{f}
\end{eqnarray}
Thus, from Eqs. (\ref{T-general})-(\ref{f}) we get,
\begin{eqnarray}
V_{l=0}(k) = -\frac{4\pi \hbar^2}{m} \frac{\tan[\delta(k)]}{k}.
\label{V}
\end{eqnarray}
So by using the expansion given by Eq. (\ref{fr}) into Eq. (\ref{V}), and taking
into account the Taylor expansion of  $V_{l=0}(k)$ with respect to $k$, such 
that, $V_{l=0}(k)=g^{(0)} + g^{(2)} k^2 + ...$, we obtain the following 
interacting $3d$ Bose-Bose intra-species (inter-species) coupling constants
\begin{eqnarray}
g^{(0)} = \frac{4\pi\hbar^2}{m}a,\quad \quad \quad \quad
g^{(0)}_{12} = \frac{4\pi\hbar^2}{m}a_{12},
\label{couplings3d}
\end{eqnarray}
and
\begin{eqnarray}
g^{(2)} = \frac{2\pi\hbar^2}{m}a^2r, \quad \quad \quad \quad
g^{(2)}_{12} = \frac{2\pi\hbar^2}{m}a^2_{12}r_{12}.
\label{couplings3d-fr}
\end{eqnarray}
Expressions in Eq. (\ref{couplings3d}) are well known for dilute and ultracold
atomic gases.
This kind of couplings arise when it is worthwhile replacing the two-body 
interaction potential between atoms with a pseudo-potential.
The most usual, and simple, scheme consists in using the zero-range Fermi
pseudo-potential.
Nevertheless, an improvement of the previous contact or zero-range approximation
can be achieved by replacing the interaction potential with a pseudo-potential
\cite{Nonuniversal1,Nonuniversal2}.
In such a case it is possible to obtain the expressions in Eq.
(\ref{couplings3d-fr}). 

It is remarkable to mention that applicability of expansion (\ref{fr}) holds 
under the condition $r \ll a$.
In addition a physically reasonable assumption to an expansion for $r \gg  a$ is
given by \cite{Adhikari}
\begin{eqnarray}
\frac{1}{k}\tan[\delta (k)] = - a + \frac{1}{6}r^3k^2 +...,
\label{fr-2}
\end{eqnarray}
which gives a good description of low-energy scattering processes.
Instead Eq. (\ref{Tk}) leads to the new nonuniversal couplings
\begin{eqnarray}
g^{(2)} = -\frac{2\pi\hbar^2}{3m}r^3, \quad \quad \quad \quad
g^{(2)}_{12} = -\frac{2\pi\hbar^2}{3m}r^3_{12},
\label{couplings3d-fr-2}
\end{eqnarray}
The coupling constants for zero-range interactions keep as in 
Eq. (\ref{couplings3d}).

From the symmetric grand-potential in Eq. (\ref{grand-symmetric}) and using the 
coupling for zero-range interactions in Eq. (\ref{couplings3d}), the energy per 
particle in the mean-field approximation can be read as
\begin{eqnarray}
\frac{E_0}{N}=\frac{1}{4}\Big(1+\frac{a_{12}}{a}\Big)g^{(0)}n.
\label{E0}
\end{eqnarray}
On the other hand, from Eq. (\ref{grand-pot}), with $d=3$, and performing
dimensional regularization we obtain the grand-potential
\cite{1D-3D-mixtures,Regularization}
\begin{eqnarray}
\frac{\Omega_g}{L^3}=\frac{8}{15\pi^2} \Big(\frac{m}{\hbar^2}\Big)^{3/2}
\frac{\mu^{5/2}}{(1+\delta_+\mu)^2}
+ \frac{8}{15\pi^2} \Big(\frac{m}{\hbar^2}\Big)^{3/2}
\frac{\lambda^{5/2}\mu^{5/2}}{(1+\delta_-\mu)^2}.
\end{eqnarray}
From this grand potential we also obtain the Lee-Huang-Yang-like (LHY-like) 
correction to the energy per particle.
Therefore the total energy per particle at zero-temperature of $3d$ Bose-Bose
mixtures including nonuniversal corrections to the interactions can be read as
\begin{eqnarray}
\frac{E_{NUC}}{N} &=& (a+a_{12})\frac{\pi \hbar^2}{m} n
\nonumber \\
&+&\frac{32\sqrt{2\pi}}{15} \frac{\hbar^2}{m}
\frac{(a+a_{12})^{5/2}n^{3/2}}{\big[1-\frac{4}{3}\pi\big(r^3+r^{3}_{12}
\big)n\big]^2}
\nonumber \\
&+&\frac{32\sqrt{2\pi}}{15} \frac{\hbar^2}{m}
\frac{(a-a_{12})^{5/2}n^{3/2}}{\big[1-\frac{4}{3}\pi\big(r^3-r^{3}_{12}
\big)n\big]^2},
\label{Total-energy-FR}
\end{eqnarray}
we have choose the couplings given in Eq. (\ref{couplings3d-fr-2}). Instead for
couplings in Eq. (\ref{couplings3d-fr}) we can use
\begin{eqnarray}
r^3 \pm r^{3}_{12}
\rightarrow -3(a^2r \pm a^{2}_{12}r_{12}).
\label{change}
\end{eqnarray}

In absence of the nonuniversal effects of the interactions we recover the result
obtained in Refs. \cite{Petrov1}, and \cite{1D-3D-mixtures}, such that
\begin{eqnarray}
\frac{E}{N} = (a+a_{12})\frac{\pi \hbar^2}{m} n
+\frac{32\sqrt{2\pi}}{15} \frac{\hbar^2}{m}
[(a+a_{12})^{5/2}+(a-a_{12})^{5/2}]n^{3/2}.
\label{Energy}
\end{eqnarray}
In presence or in absence of the nonuniversal corrections to the interactions we
can see that for attractive inter-species interactions ($a_{12}<0$) slightly 
larger than repulsive intra-species interactions ($a>0$), such that,
$|a_{12}|/|a| \sim -1^-$, the energy per particle becomes complex, rendering the
mixture unstable \cite{Petrov1}.
However, if we consider the approximation $-a_{12}=a$ in the fluctuations only,
as it was done by Petrov in Ref. \cite{Petrov1}, the energy per particle keeps
real, and Eq. (\ref{Energy}) reduces to the free droplet phase with energy per
particle
\begin{eqnarray}
\frac{E_{D}}{N} = -|a - a_{12}|\frac{\pi\hbar^2}{m} n
+\frac{256\sqrt{\pi}}{15} \frac{\hbar^2}{m}
a^{5/2} n^{3/2}
\label{Energy-P}.
\end{eqnarray}

It is worth mentioning that the term $(a+a_{12})^{5/2}+(a-a_{12})^{5/2}$ in the 
fluctuations contribution of Eq. (\ref{Energy}) is symmetric under 
$a_{12} \rightarrow -a_{12}$. Therefore the same droplet can arise in another
equivalent situation where $a_{12}=a$ in the fluctuations only 
\cite{Petrov1,Montecarlo-Bose-Bose1}.
So providing that $a_{12}= a$ or $-a_{12}= a$ just in the Gaussian corrections to
the total energy per particle in Eq. (\ref{Total-energy-FR}), the system exhibits
a liquidlike phase where the $3d$ droplet with nonuniversal corrections to the
interactions (DNUC) emerges with energy per particle
\begin{eqnarray}
\frac{E^{\pm}_{DNUC
}}{N} = -|a - a_{12}|\frac{\pi\hbar^2}{m} n
+\frac{256\sqrt{\pi}}{15} \frac{\hbar^2}{m}
\frac{a^{5/2} n^{3/2} }{\big[1-\frac{4}{3}\pi \big(r^3\pm r^3_{12}\big)n\big]^2}.
\label{FRD}
\end{eqnarray}

\begin{figure}[t]
\begin{center}
\includegraphics[height=7cm, width=8cm, clip]{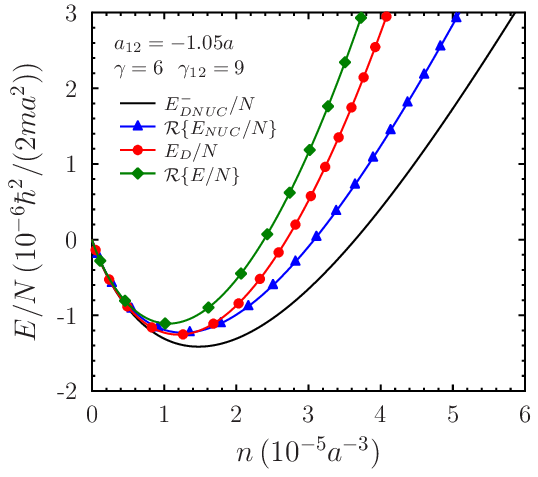}
\end{center}
\vskip -.8cm
\caption{Three-dimensional energy per particle of Bose-Bose gases with 
nonuniversal corrections to the interactions as function of the density.
We consider attractive inter-species interactions such that $a_{12}=-1.05a$.
We use $\hbar^2/(2ma^2)$ as the energy unit, and we set
$\gamma = r/a =6$, and $\gamma_{12} = r_{12}/a=9$.
The energy per particle of droplets with nonuniversal corrections to the 
interactions $E^{-}_{DNUC}/N$ is given by Eq. (\ref{FRD}) (solid black line).
The real contribution of Eq. (\ref{Total-energy-FR}) is 
$\mathcal{R}\{E_{NUC}/N\}$ with $a<-a_{12}$,  (blue triangles).
We also include the usual droplet given in Eq. (\ref{Energy-P}), $E_D/N$, 
(red circles). Real contribution of Eq. (\ref{Energy}), $\mathcal{R}\{E/N\}$, 
(green diamonds), is also included.
}
\label{F1}
\end{figure}

We use the couplings in Eq. (\ref{couplings3d-fr-2}) considering that
in absence of the fluctuations the attractive effect of the interactions, such 
that $-a_{12}>a$, causes the condensate to collapse therefore the strong
attraction lead us to consider the nonuniversal corrections to the attractive 
interactions such that $r_{12}\gg a_{12}$, as a relevant contribution. 
However we must keep in mind that although attraction is greater than repulsion,
the repulsion is very close to the attraction and in turn such a repulsion 
encoded in $a$ controls the fluctuations. 
In this sense, and in order to have a more complete overview of the effects of 
the nonuniversal corrections to the interactions we also include the nonuniversal
correction to the repulsion such that $r \gg a$.
Nevertheless neglecting the nonuniversal corrections to the repulsive 
interactions, and using the nonuniversal contribution of the attractive 
interactions as a fitting parameter we obtain a good agreement between our 
theoretical predictions and some diffusion Monte Carlo calculations (DMC) 
\cite{Montecarlo-Bose-Bose1}, as we will see later.
Experimental values of the s-wave scattering length and the nonuniversal 
corrections to the interactions as a function of the external magnetic field in 
ultracold $^{39}$K atoms show that it is possible to satisfy the condition 
$r \gg  a$ \cite{Montecarlo-Bose-Bose2,39K}.

In Fig. \ref{F1} we plot the energy per particle given by Eqs. 
(\ref{Total-energy-FR}), and (\ref{Energy})-(\ref{FRD}). We consider
$\hbar^2/(2ma^2)$ as the energy unit, and we define $\gamma = r/a$, and 
$\gamma_{12} = r_{12}/a$.
The energies are considered for an experimentally relevant density in the
conventional droplet regime, $na^3 \sim 10^{-5}$ 
\cite{Exper-droplet1,Exper-droplet2}.
We choose the negative sign for the DNUC in Eq. (\ref{FRD}).
We set $a<-a_{12}$ in the fluctuations of Eq. (\ref{Total-energy-FR}), and
$\mathcal{R}\{E_{NUC}/N\}$ corresponds to real part of (\ref{Total-energy-FR}).
As a physically reasonable assumption we neglect the imaginary term providing
$\mathcal{I}\{E_{NUC}/N\}\ll \mathcal{R}\{E_{NUC}/N\}$.
In a similar way we have $\mathcal{R}\{E/N\}$ using Eq. (\ref{Energy}).
At very low densities $n \lesssim 5\times 10^{-6}a^{-3}$ the nonuniversal
contribution to the energy in Eqs. (\ref{Total-energy-FR}) and (\ref{FRD}) could
be neglected, reproducing the energies in Eqs. (\ref{Energy}) and 
(\ref{Energy-P}) respectively.
In turn the energies in Eqs. (\ref{Energy}) and (\ref{Energy-P}) produce almost 
the same result, as it is expected.
However, for values greater than this density we can see a clear difference
between models predictions.
For both the DNUC and the real contribution of the energy per particle including
the nonuniversal effects of the interactions the density of particles is greater 
than their counterparts without the enhanced nonuniversal effects.
This behavior can be understood in the following way, if 
$\gamma^3_{12}-\gamma^3 > 0$ the nonuniversal correction to the attraction
exceeds that of repulsion.
Thus the effective contribution given by the nonuniversal effects to the 
interactions allows an increasing in the attraction which in turn allows that 
equilibrium density increases.
For large values of $\gamma^3_{12}-\gamma^3 > 0$ inelastic collisions decreases
the number of atoms that can be accommodated in a self-bound state and we no
longer find a minimum in the energy per particle.
The DNUC undergoes instability and it disappears. There is not a stable 
self-bound state.
Otherwise, when the nonuniversal correction to the repulsion exceeds that of 
attraction, holding the condition $\gamma^3_{12} - \gamma^3 < 0$, the total 
effective repulsion causes decreasing in the number of atoms forming the DNUC.
So the equilibrium density and the associated energy are lower than their
counterparts without the nonuniversal effects.
Using the couplings of Eq. (\ref{couplings3d-fr}) we have 
$\gamma, \gamma_{12} \ll 1$ and their effects on the energies per particle
of Eqs. (\ref{Total-energy-FR}) and (\ref{FRD}) are not perceptible.
It is remarkable that the behavior presented in Fig. \ref{F1} is similar to one 
obtained for usual droplets in numerical calculations
\cite{Numeric-drop,Montecarlo-Bose-Bose1, Montecarlo-Bose-Bose2}. 

\begin{figure}[t]
\begin{center}
\includegraphics[height=5cm, width=12cm, clip]{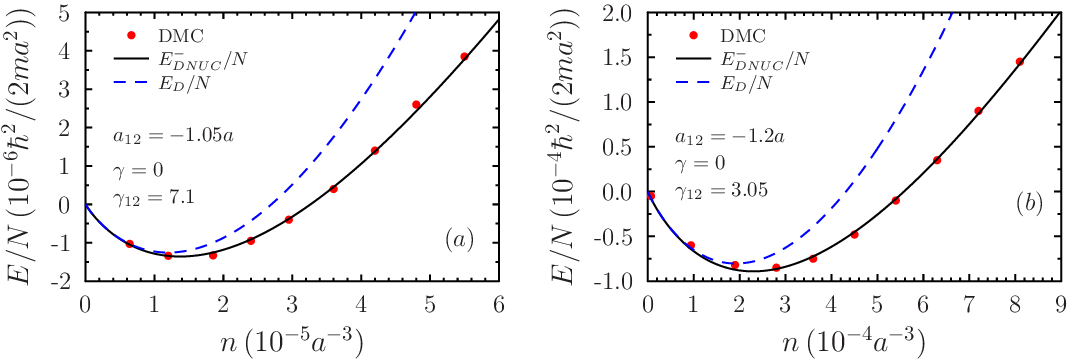}
\end{center}
\vskip -.6cm
\caption{Three-dimensional energy per particle of Bose-Bose gases with 
nonuniversal corrections to the interactions as function of the density.
The energy per particle of DNUC, $E^{-}_{DNUC}/N$ is given by Eq. (\ref{FRD})
(solid black line).
The usual droplet given in Eq. (\ref{Energy-P}), $E_D/N$, (dashed blue line). 
The diffusion Monte Carlo (DMC) results are given from  
\cite{Montecarlo-Bose-Bose1}  (red dots).
We use $\hbar^2/(2ma^2)$ as the energy unit. (a) We set the attractive 
inter-species interactions such that $a_{12}=-1.05a$. The nonuniversal 
corrections for the repulsion and for the attraction are $\gamma =0$ and
$\gamma_{12} =7.1$, respectively.
(b) we consider $a_{12}=-1.2a$, $\gamma =0$ and $\gamma_{12} =3.05$.
}
\label{F2}
\end{figure}

In Fig. \ref{F2} we compare the energy per particle of DNUC, given by 
Eq. (\ref{FRD}) with some DMC results given in \cite{Montecarlo-Bose-Bose1}.
We consider two different attractive inter-species interactions such that 
$a_{12}=-1.05a$, and $a_{12}=-1.2a$. Taking into account that the attraction is 
leading  we set the nonuniversal effects to the repulsive interactions as 
$\gamma =0$.
The nonuniversal effects to the attractive interactions are chosen as fitting 
parameters with $\gamma_{12} =7.1$ and $\gamma_{12} =3.05$.
We also include the energy per particle for the usual droplet given in 
Eq. (\ref{Energy-P}).
In particular the minimum energy $n_0$ predicted by Eq. (\ref{Energy-P}) is 
$\sim 1.198$, and by means of Eq. (\ref{FRD}) we found $\sim 1.476$, so we have 
$n_{0,DNUC} \sim 1.232n_{0,D}$. We can also see that if $a\sim 100$\AA $\,$
we obtain densities with experimental relevance. Thus  
$n \sim 10^{13}\text{atoms}/\text{cm}^3$ and
$n \sim 10^{14}\text{atoms}/\text{cm}^3$ in Fig. \ref{F2}(a) and 
Fig. \ref{F2}(b) respectively. 

\begin{figure}[t]
\begin{center}
\includegraphics[height=8cm, width=12cm, clip]{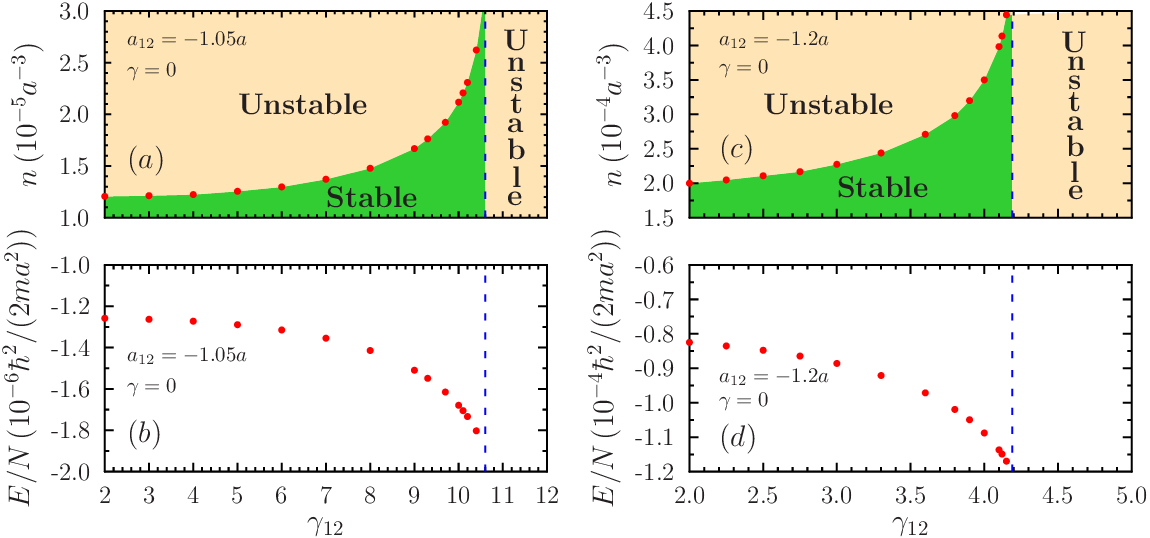}
\hspace{0.2cm}
\end{center}
\vskip -.6cm
\caption{Phase diagram and energy per particle for DNUC as function of the 
nonuniversal effects of the attractive interactions $\gamma_{12}$. The red dots 
are obtained from  Eq. (\ref{FRD}) as the minimum density or minimum energy per
particle for a stable DNUC formation.
The nonuniversal correction to the repulsion is fixed as $\gamma =0$.
We use $\hbar^2/(2ma^2)$ as the energy unit.(a)-(b) we set the attractive 
inter-species interactions such that $a_{12}=-1.05a$. 
(c)-(d) we consider $a_{12}=-1.2a$. The dashed blue line represents the 
threshold value for $\gamma_{12}$.
}
\label{F3}
\end{figure}

\begin{figure}[t]
\begin{center}
\includegraphics[height=5cm, width=12cm, clip]{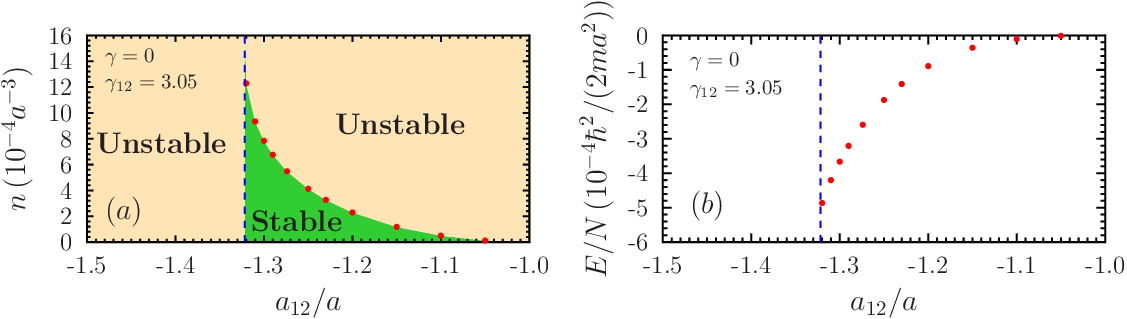}
\end{center}
\vskip -.6cm
\caption{Phase diagram and energy per particle for DNUC as function of the 
dimensionless attractive scattering length $a_{12}/a$. The red dots are obtained 
from  Eq. (\ref{FRD}) as the minimum density or minimum energy per particle for a
stable DNUC formation.
The nonuniversal correction for the repulsion is fixed as $\gamma =0$.
We use $\hbar^2/(2ma^2)$ as the energy unit. We set the nonuniversal correction
to the attractions as $\gamma_{12}=3.05$. 
The dashed blue line represents the threshold value for $a_{12}/a$.
}
\label{F4}
\end{figure}

By using Eq. (\ref{FRD}) we plot the phase diagram and energy per particle for
DNUC as function of the  nonuniversal effects of the attractive interactions 
$\gamma_{12}$ in Fig. \ref{F3}. 
The nonuniversal correction to the repulsion is fixed as $\gamma =0$.
We use $\hbar^2/(2ma^2)$ as the energy unit. In  Fig. \ref{F3}(a)-(b) we set the
attractive inter-species interactions such that $a_{12}=-1.05a$. 
In Fig. \ref{F3}(c)-(d) we consider $a_{12}=-1.2a$. 
The blue line represents the threshold values of $\gamma_{12}$ with 
$\gamma_{12} \sim 10.6$  and $\gamma_{12} \sim  4.19$,
for $a_{12}=-1.05a$, and $a_{12}=-1.2a$, respectively.
In Fig. \ref{F3} we show how increasing the nonuniversal correction to the 
attractions, the density increases, the DNUC is more attractive and eventually 
this becomes unstable, as it was explained above.
We can also see how the threshold for the instability is smaller as the 
attractive scattering length increases, as it is expected. This can be understood
due to the fact that the leading contribution is attractive.
In Fig. \ref{F4} we also plot the phase diagram and energy per particle for DNUC 
as function of the dimensionless attractive scattering length
$a_{12}/a$ with fixed values of $\gamma$ and $\gamma_{12}$. This behavior 
is equivalent to that presented in the Fig. \ref{F3} and it is an obvious 
consequence of the DNUC instability when the leading contribution, the 
attraction, increases. In other words for values greater than 
$a_{12}/a\sim -1.3217$ the DNUC collapses.

Now, we present an improved $3d$ Gross-Pitaevskii equation (GPE) to describe the
phase of droplets with nonuniversal corrections to the interactions in free
space
\begin{eqnarray}
\text{i}\hbar \frac{\partial \phi}{\partial t}
&=& \Big[-\frac{\hbar^2}{2m} \nabla^2
- \frac{2\pi\hbar^2}{m}|a_{12}-a||\phi|^2
+ \frac{\pi\hbar^2}{3m}(r^3 + r^3_{12})\nabla^2 |\phi|^2
\nonumber \\
&+&
\frac{128\sqrt{\pi}}{3} \frac{\hbar^2}{m} a|\phi|^2 \sqrt{|\phi|^2 a^3}
F^{\pm}(r,r_{12},|\phi|^2) \Big]\phi,
\label{3D+FR+GPE}
\end{eqnarray}
where
\begin{eqnarray}
F^{\pm}(r,r_{12},|\phi|^2)  = \frac{1-\frac{4\pi}{15}
\big(r^3 \pm r^3_{12}\big)|\phi|^2}
{\big[1- \frac{4\pi}{3} \big(r^3 \pm r^3_{12}\big)|\phi|^2\big]^3}.
\end{eqnarray}
The first two terms correspond to the usual attractive mean-field approximation 
in Bose-Bose gases.
The third term describes the nonuniversal corrections to the interactions at 
level of mean-field approximation.
The last contribution takes into account improved fluctuation effects including 
the nonuniversal corrections to the interactions where it was used 
$a=\pm a_{12}$.
By increasing the density could be possible to find an instability associated to
$F^{\pm}$, for both $a \gg r$ and $a \ll r$.  
For $a \ll r$ the instability is present for
$|\phi|^2 = 3/[4\pi(r^3 \pm r^3_{12})]$, with $r>r_{12}$ choosing the negative 
sign. 
For $a \gg r$ the DNUC becomes unstable for 
$|\phi|^2 = 1/[4\pi(a^2_{12}r_{12} - a^2r)]$, with $a^2_{12}r_{12}>a^2r$.
Since $r^3+r^3_{12}>0$, the nonuniversal corrections to the interactions are 
repulsive at mean-field level. 
At first glance, this would allow controlling the number of atoms into the 
self-bound liquidlike state. 
Increasing the strength of the repulsion beyond a critical value the atoms could 
be spreading out decreasing on the number of atoms necessary to obtain a stable 
DNUC. This could be helpful for fitting with experimental or numerical evidence.
It is also remarkable and interesting to consider only the mean-field 
approximation,
without fluctuations. The attraction-repulsion balance could allow a few atoms to
be accommodated into a stable bound-state formation, but not including Gaussian 
corrections.
Although the nonuniversal corrections to the interactions are not leading, these 
could in principle play a similar role as the LHY contribution in a conventional 
droplet.
In a different scenario this equation could be useful to study how solitons and
vortices are affected by such a nonuniversal corrections to the interactions.
Finally, taking into account the improved GPE with the change given by 
Eq. (\ref{change}) it could be interesting even in a context other than droplets.

\subsection{Three-dimensional finite-temperature effects}

We briefly describe some aspects regarding finite-temperature corrections to the
equation of state of $3d$ Bose-Bose mixtures with nonuniversal corrections to the
interactions. 
In the continuum limit the $3d$ one-loop correction to the grand potential in
Eq. (\ref{Grand-T}) is written as \cite{1D-3D-mixtures},
\begin{eqnarray}
\frac{\Omega_{g}^{(T)}}{L^3}
= -\frac{1}{6\pi^2} \sum_{\pm}\int_0^{\infty} dE_{\pm}(k_{\pm})
\frac{k^{3}_{\pm}}{e^{\beta E_{\pm}(k)}-1},
\end{eqnarray}
and Bogoliubov spectra in Eq. (\ref{Bogol}) lead to
\begin{eqnarray}
k^2_{+}  = \frac{2m}{\hbar^2} \frac{\mu}{1+\alpha_+\mu}
\bigg(\sqrt{1+\frac{(1+\alpha_+\mu)}{\mu^2 }E^2_+} -1 \bigg),
\label{k+}
\end{eqnarray}
and
\begin{eqnarray}
k^2_{-}  = \frac{2m}{\hbar^2} \frac{\lambda\mu}{1+\alpha_-\mu}
\bigg(\sqrt{1+\frac{(1+\alpha_-\mu)}{\lambda^2\mu^2 }E^2_-} -1 \bigg),
\label{k-}
\end{eqnarray}
with $\lambda$ given in Eq. (\ref{constants}).
Now, after introducing the variables $x_{\pm}= \beta E_{\pm}$, and expanding
$k_{\pm}$ at low temperature $k_B T\ll |\mu|$ \cite{Pit-Strin,Peth-Smith}, we get
\begin{eqnarray}
\frac{\Omega^{(T)}_g}{L^3}
&=& - \frac{\pi^2}{90}\Big(\frac{m}{\hbar^2}\Big)^{3/2}
\frac{(k_B T)^4}{\mu ^{3/2}} \Big[ 1- \frac{5\pi^2}{7}
\frac{(1+\alpha_+ \mu)}{\mu^2}(k_B T)^2\Big]
\nonumber \\
&-&\frac{\pi^2}{90}\Big(\frac{m}{\hbar^2}\Big)^{3/2}
\frac{(k_B T)^4}{\lambda^{3/2}\mu ^{3/2}} \Big[ 1- \frac{5\pi^2}{7}
\frac{(1+\alpha_- \mu)}{\lambda^2\mu^2}(k_B T)^2\Big].
\end{eqnarray}
From which it follows the thermal energy per particle
\begin{eqnarray}
\frac{E^{(T)}_g}{N}
&=& \frac{\sqrt{2\pi}}{240} \Big(\frac{m}{\hbar^2}\Big)^{3}
\frac{(k_B T)^4}{(a+a_{12})^{3/2}n^{5/2}} (1+h^{+}_{NUC})
\nonumber \\
&+& \frac{\sqrt{2\pi}}{240} \Big(\frac{m}{\hbar^2}\Big)^{3}
\frac{(k_B T)^4}{(a-a_{12})^{3/2}n^{5/2}}(1+h^{-}_{NUC}),
\label{Energy-T}
\end{eqnarray}
with
\begin{eqnarray}
h^{\pm}_{NUC} &=& - \frac{5}{84} \Big(\frac{m}{\hbar^2}\Big)^2
\frac{(k_B T)^2}{(a \pm a_{12})^2 n^2}
\Big[ 3-\frac{20\pi}{3}(r^3 \pm r^3_{12})n\Big],
\end{eqnarray}
where we used the coupling constants from expression (\ref{couplings3d-fr-2}).
It is remarkable that unlike the zero-temperature corrections the above thermal 
improvement is divergent provided $a=\pm a_ {12}$ and the mixture undergoes a 
thermally instability.
In a similar way for attractive interspecies interactions lightly larger than
repulsive intra-species interaction, Eq. (\ref{Energy-T}) becomes complex for
both DNUC and usual droplets ($r=r_{12}=0$).
However, compared to corrections in  Eq. (\ref{Total-energy-FR}), in this case 
the imaginary contribution is leading.
Therefore both the DNUC and the usual droplets are thermal-induced unstables.
We also highlight that the nonuniversal corrections appear only up at order 
$T^6$ and, hence at very low temperatures these could be neglected.

\section{Two-dimensional model}
\label{Two-dimensional model}

We focus on two-dimensional Bose-Bose gases. Two-dimensional scattering theory 
indicates that the expansion of the phase shift $\delta(k)$, can be read as
\cite{2D}
\begin{eqnarray}
\cot[\delta(k)]= \frac{2}{\pi} \ln\Big(\frac{k}{2}a e^\gamma \Big)
+\frac{1}{2}rk^2 + ...,
\label{Expansion-2D}
\end{eqnarray}
with the s-wave scattering length $a$, the s-wave effective range of interaction 
$r$, the condition $a\gg r$, and $\gamma = 0.5572...$ the Euler-Mascheroni 
constant.
The $2d$ coefficient $C(k)$ in the s-wave transition element given in Eq.
(\ref{T-general}) is written as \cite{On-shell}
\begin{eqnarray}
C(k) = \frac{m}{2\pi\hbar^2}\ln \Big(\frac{k}{\sqrt{\epsilon_c}}\Big)
- \frac{m}{4\hbar^2}\text{i},
\label{C-2D}
\end{eqnarray}
with the low-energy cutoff $\epsilon_c = \hbar^2 k^2/m$.
In $2d$ scattering theory the s-wave phase shift $\delta(k)$ and the s-wave 
transition element $T_{l=0}(k)$ are related by means of \cite{Scattering-2D}
\begin{eqnarray}
T_{l=0}(k) = -\frac{4\pi\hbar^2}{m} \frac{1}{\cot[\delta(k)]-\text{i}}.
\label{T-2D}
\end{eqnarray}
So, inserting Eq. (\ref{C-2D}) into Eq. (\ref{T-general}), and comparing with 
(\ref{T-2D}) we get,
\begin{eqnarray}
V_{l=0}(k) = -\frac{4\pi \hbar^2}{m} \frac{1}{\cot[\delta(k)]
- \frac{2}{\pi}\ln \big(\frac{k}{\sqrt\epsilon_c}\big)}.
\label{V-2D}
\end{eqnarray}
Inserting Eq. (\ref{Expansion-2D}) into Eq. (\ref{V-2D}) and taking into account
the Taylor expansion at low momentum of $V_{l=0}(k)$ where
$V_{l=0}(k)=g^{(0)} + g^{(2)} k^2 + ...$, we can obtain the $2d$ intra-species
interacting coupling constants
\begin{eqnarray}
g^{(0)}=\frac{4\pi\hbar^2}{m}\frac{1}{\ln
\big(\frac{4 e^{-2\gamma}}{a^2\epsilon_c}\big)},
\label{g0_2d}
\end{eqnarray}
and
\begin{eqnarray}
g^{(2)}=-\frac{2\pi^2\hbar^2}{m}\frac{r^2}
{\ln^2\big(\frac{4 e^{-2\gamma}}{a^2\epsilon_c}\big)}.
\label{g2_2d}
\end{eqnarray}
In a similar way the inter-species coupling constants are
\begin{eqnarray}
g^{(0)}_{12}=\frac{4\pi\hbar^2}{m}\frac{1}{\ln\big(\frac{4
e^{-2\gamma}}{a_{12}^2\epsilon_c}\big)},
\label{g0_12_2d}
\end{eqnarray}
and,
\begin{eqnarray}
g^{(2)}_{12}=-\frac{2\pi^2\hbar^2}{m}\frac{r^2_{12}}
{\ln^2\big(\frac{4 e^{-2\gamma}}{a^2_{12}\epsilon_c}\big)}.
\label{g2_12_2d}
\end{eqnarray}

From symmetric grand-potential in Eq. (\ref{grand-symmetric}) and the coupling 
constants of zero-range interactions in Eqs. (\ref{g0_2d}) and (\ref{g0_12_2d}),
the $2d$ mean-field energy per particle is given by
\begin{eqnarray}
\frac{E}{N} = \frac{1}{4} \Big(1+ \frac{g^{(0)}_{12}}{g^{(0)}} \Big)g^{(0)}n.
\end{eqnarray}
In Eq. (\ref{grand-pot}) with $d = 2$ and performing dimensional regularization
in the modified minimal subtraction scheme $\mathrm{\overline{MS}}$-scheme 
\cite{Rabi-E}, lead us to the following grand-potential including nonuniversal
corrections to the interactions
\begin{eqnarray}
\frac{\Omega_g}{L^2} &=&
-\frac{m}{8\pi\hbar^2}\frac{\mu^2}{(1+\delta_+\mu)^{3/2}}
\ln\Big[ \frac{\epsilon_c}{\sqrt{e}} \frac{(1+\delta_+\mu)}{\mu}\Big]
\nonumber \\
&-& \frac{m}{8\pi\hbar^2}\frac{\lambda^2\mu^2}{(1+\delta_-\mu)^{3/2}}
\ln\Big[ \frac{\epsilon_c}{\sqrt{e}} \frac{(1+\delta_-\mu)}{\lambda\mu}\Big].
\end{eqnarray}
The $2d$ energy per particle of Bose-Bose mixtures could be obtained through the 
conventional thermodynamic relations in the grand-canonical ensemble.
However we are interested and we will focus on the case of weakly attractive
inter- and weakly repulsive intra-species interactions, where
$a_{12}^{-1}\ll \sqrt{n}\ll a^{-1}$ \cite{Petrov2,Rabi-E}. 
For simplicity, we set $m =\hbar = 1$.
In order to obtain the energy per particle in this regime we introduce a new
energy cutoff $\tilde{\epsilon}_c$, such that, 
$\tilde{\epsilon}_c = 4 e^{-2\gamma}/(aa_{12})$ \cite{Petrov2,Rabi-E}.
Provided that $\tilde{\epsilon}_c/\epsilon_c$ is not exponentially large we can
use $\epsilon_c \rightarrow \tilde{\epsilon_c}$ as a good assumption from which 
we get the following coupling constants
\begin{eqnarray}
g^{(0)} = \frac{4\pi}{\ln(a_{12}/a)}, \quad\quad\quad
g^{(0)}_{12} = \frac{4\pi}{\ln(a/a_{12})},
\label{g0-2d-simplif}
\end{eqnarray}
where $g^{(0)}  = -g^{(0)}_{12}$. In a similar way, the coupling constants with 
nonuniversal corrections to the interactions are
\begin{eqnarray}
g^ {(2)} = -\frac{2\pi^2 r^2}{\ln^2(a_{12}/a)}, \quad\quad\quad
g^ {(2)}_{12} = -\frac{2\pi^2 r^2_{12}}{\ln^2(a_{12}/a)},
\end{eqnarray}
where  $g^{(2)}_{12}/g^{(2)} = (r_{12}/r)^2$.
So we get the energy per particle of $2d$ DNUC as
\begin{eqnarray}
\frac{E_{DNUC}}{N} = \frac{1}{8\pi} \frac{[g^{(0)}]^2 n}{
\big[1+2g^{(2)}\big[1 - (\frac{r_{12}}{r})^2\big]n\big]^{3/2}}
\ln \Big[\frac{e^{2\gamma +1/2}}{4} \frac{aa_{12}g^{(0)}n}
{1+2g^{(2)}\big[1 - \big(\frac{r_{12}}{r}\big)^2\big]n} \Big].
\label{NUC-E-2d}
\end{eqnarray}

In absence of the nonuniversal effects to the interactions or in the symmetric 
case $r=r_{12}$, the energy per particle in Eq. \ref{NUC-E-2d} describes the 
conventional $2d$ droplet phase \cite{Petrov2}, where
\begin{eqnarray}
\frac{E}{N} = \frac{2\pi n}{\ln^2(a_{12}/a)}
\ln \Big(\frac{n}{e n_0} \Big),
\label{E-2d}
\end{eqnarray}
with $n_0$ the equilibrium density
\begin{eqnarray}
n_ 0 = \frac{1}{\pi} e^{-2\gamma - 3/2} \frac{\ln(a_{12}/a)}{a a_{12}}.
\label{equilib-density}
\end{eqnarray}
From Eqs. \ref{E-2d} and \ref{equilib-density} we can write Eq. \ref{NUC-E-2d} as 
\begin{eqnarray}
\frac{E_{DNUC}}{N} = \frac{2\pi\, n}{\ln^2(a_{12}/a)}\frac{1}{h_{NUC}^{3/2}}
\ln \Big(\frac{n}{en_0} \frac{1}{h_{NUC}} \Big),
\label{FR-E-2d}
\end{eqnarray}
with $h_{NUC} = 1+ f_{NUC}|\phi|^2$, and $f_{NUC} = 2 g^{(2)}[1- (r_{12}/r)^2]$.

\begin{figure}[t]
\begin{center}
\includegraphics[height=7cm, width=8cm,clip]{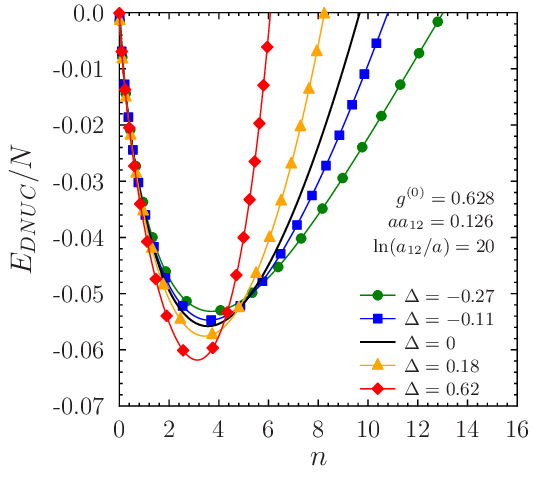}
\end{center}
\vskip -.8cm
\caption{Two-dimensional energy per particle of droplets with nonuniversal
corrections to the interactions as function of the density, 
Eq. (\ref{FR-E-2d}).
We fix $g^{(0)}= 0.628$, $aa_{12}=0.126$ and $\ln(a_{12}/a)= 20$, and we define
$\Delta \equiv 1- (r_{12}/r)^2$.
We include $\Delta = -0.27$ (green circles), $\Delta = -0.11$ (blue squares), 
$\Delta = 0.18$ (orange triangles) and $\Delta = 0.62$ (red diamonds).
$\Delta = 0$ (solid black line) corresponds to absence of nonuniversal
improvements to the interactions given in Ref. \cite{Petrov2}.
}
\label{F5}
\end{figure}

In Fig. \ref{F5} we plot the energy per particle from Eq. (\ref{FR-E-2d}) as 
density function.
We set the values $g^{(0)}= 0.628$, $aa_{12}=0.126$ and $\ln(a_{12}/a)= 20$.
We define $\Delta \equiv 1- (r_{12}/r)^2$, and we consider different values and 
change of sign for this parameter.
The case $\Delta = 0$ corresponds to results in absence of nonuniversal 
corrections to the interactions obtained in Ref. \cite{Petrov2}.
For $\Delta < 0$ the nonuniversal correction to the attraction exceeds that of 
repulsion.
However the equilibrium density is only slightly larger than in absence of
nonuniversal effects to the interactions. 
If nonuniversal correction to the repulsion exceeds that of attraction, it holds
the condition $\Delta > 0$, and the equilibrium density is only slightly lower 
than in absence of nonuniversal effects of the interactions. 
A possible explanation for not observing significant changes could be associated
with the use of coupling constants. In such a case we consider $a\gg r$, but 
taking into account the perturbative nature of the self-bound liquidlike phase 
could be suitable consider coupling constants associated to $a\ll r$.

Now, we introduce the improved $2d$ GPE to describe free DNUC
\begin{eqnarray}
\text{i} \frac{\partial \phi}{\partial t}
&=& \Big[-\frac{1}{2} \nabla^2
+ \frac{2\pi^2}{\ln^2(a_{12}/a)} (r^2+r^2_{12})\nabla^2 |\phi|^2
\nonumber \\
&+& \frac{4\pi}{\ln^2(a_{12}/a)} \frac{|\phi|^2}{h_{NUC}^{5/2}}
\Big[ \big( 1 + \frac{1}{4}f_{NUC}|\phi|^2 \big)
\ln\Big(\frac{|\phi|^2}{e n_0}\frac{1}{h_{NUC}}\Big)+\frac{1}{2}\Big]\Big]
\phi, 
\label{2D+FR+GPE}
\end{eqnarray}
where $n_0$ is given in Eq. (\ref{equilib-density}).
The first line corresponds to the mean-field approximation.
The second line corresponds to the fluctuations including the nonuniversal
corrections to the interactions.
At first glance, given Eq. (\ref{g0-2d-simplif}), the coupling of the
nonuniversal corrections to the interactions provides a repulsive contribution
at the mean-field level in Eq. (\ref{2D+FR+GPE}).
However, the dynamics of this $2d$ improved GPE is much more complex. We are
including the dependence on the nonuniversal contribution to the interactions to 
the already known logarithmic dependence.
In particular, under the physically reasonable assumption which nonuniversal
contributions to the interactions are neglected in the fluctuation sector a 
simplified version of the previous $2d$ improved GPE is read as
\begin{eqnarray}
\text{i} \frac{\partial \phi}{\partial t}
= \Big[-\frac{1}{2} \nabla^2
+ \frac{2\pi^2}{\ln^2(a_{12}/a)} (r^2+r^2_{12})\nabla^2 |\phi|^2
+ \frac{4\pi}{\ln^2(a_{12}/a)}|\phi|^2
\ln\big(\frac{|\phi|^2}{\sqrt{e} n_0}\big)\Big]\phi ,
\label{2D+FR+GPE-simplif}
\end{eqnarray}
This improved GPE exhibits a competition between subleading nonuniversal
corrections to the interactions at mean-field level and Gaussian fluctuations 
associated to the zero-range approximation in the interaction potential.
By means of this GPE it is possible to study how conventional $2d$ droplets and
vortex droplets \cite{Vortex-2d-drop} are affected by the nonuniversal
corrections to the interatomic interactions.

\subsection{Two-dimensional finite-temperature effects}

In the continuum limit the $2d$ one-loop correction to the grand potential in
Eq. (\ref{Grand-T}), is written as
\begin{eqnarray}
\frac{\Omega_{g}^{(T)}}{L^2}
= -\frac{1}{4\pi} \sum_{\pm}\int_0^{\infty} dE_{\pm}(k_{\pm})
\frac{k^{2}_{\pm}}{e^{\beta E_{\pm}(k)}-1},
\end{eqnarray}
with $k^2_{\pm}$ given by  Eqs. (\ref{k+}) and (\ref{k-}).
Now, after introducing the variables $x_{\pm}= \beta E_{\pm}$, and expanding
$k_{\pm}$ at low temperature $k_B T \ll |\mu|$ \cite{Pit-Strin,Peth-Smith}, it 
follows the $2d$ grand-canonical potential
\begin{eqnarray}
\frac{\Omega^{(T)}_g}{L^2} &=& - \frac{1}{4\pi}\Gamma(3)\zeta(3)
\Big( \frac{m}{\hbar^2} \Big) \frac{(k_B T)^3}{\mu}
\Big[ 1- \frac{\Gamma(5)\zeta(5)}{4\Gamma(3)\zeta(3)}
\frac{(1+\alpha_+ \mu)}{\mu^2} (k_B T)^2 \Big]
\nonumber \\
&-& \frac{1}{4\pi} \Gamma(3)\zeta(3)
\Big( \frac{m}{\hbar^2} \Big) \frac{(k_B T)^3}{\lambda \mu}
\Big[ 1- \frac{\Gamma(5)\zeta(5)}{4\Gamma(3)\zeta(3)}
\frac{(1+\alpha_- \mu)}{\lambda^2 \mu^2} (k_B T)^2 \Big],
\end{eqnarray}
with the Euler's gamma function $\Gamma(x)$ and the Riemann's zeta function
$\zeta(x)$.
Thus we get the respective thermal energy per particle as
\begin{eqnarray}
\frac{E^{(T)}_g}{N} &=& \frac{\Gamma(3)\zeta(3)}{2\pi}\Big( \frac{m}{\hbar^2}
\Big)\frac{(k_B T)^3}{[g^{(0)} + g^{(0)}_{12}]n^2}
\Big[ 1- \frac{\Gamma(5)\zeta(5)}{\Gamma(3)\zeta(3)}
\frac{(k_B T)^2}{[g^{(0)} + g^{(0)}_{12}]^2n^2} h^{+}_{NUC}  \Big]
\nonumber \\
&+& \frac{\Gamma(3)\zeta(3)}{2\pi}\Big( \frac{m}{\hbar^2} \Big)
\frac{(k_B T)^3}{[g^{(0)} - g^{(0)}_{12}]n^2}
\Big[ 1- \frac{\Gamma(5)\zeta(5)}{\Gamma(3)\zeta(3)}
\frac{(k_B T)^2}{[g^{(0)} - g^{(0)}_{12}]^2n^2} h^{-}_{NUC}  \Big],
\nonumber \\
\end{eqnarray}
with
\begin{eqnarray}
h^{\pm}_{NUC}= 1 + \frac{4m}{\hbar^2}[g^{(2)} \pm g^{(2)}_{12}]n.
\end{eqnarray}
Like in the $3d$ case, the above thermal improvements are divergent provided
$a=\pm a_ {12}$ and the Bose-Bose mixture exhibits a thermal instability.
For attractive inter-species interactions lightly larger than repulsive 
intra-species interaction, the energy per particle becomes complex with leading 
imaginary contribution.
Therefore both the $2d$ DNUC and the usual $2d$  droplets undergo thermal-induced
instability.
We also highlight that the effective nonuniversal corrections to the interactions
appear only up at order $T^5$ and, hence at very low temperatures these could be 
neglected.

\section{One-dimensional model}
\label{One-dimensional model}

In one-dimensional scattering theory the expansion of the phase shift $\delta(k)$
is given by \cite{Scatt-1d-1,Scatt-1d-2}
\begin{eqnarray}
k \tan[\delta(k)] = \frac{1}{a} + \frac{1}{2}rk^2 + ...,
\label{Expansion-1D}
\end{eqnarray}
with the s-wave scattering length $a$, the s-wave effective range of interaction
$r$, and the condition $a\gg r$.
The $1d$ coefficient $C(k)$ in the s-wave transition element given in Eq.
(\ref{T-general}) is written as \cite{On-shell}
\begin{eqnarray}
C(k) = -\frac{\text{i}}{k}\frac{m}{2 \hbar^2}.
\label{C-1D}
\end{eqnarray}
In one-dimensional scattering theory the s-wave phase shift $\delta(k)$ and the
s-wave transition element $T_{l=0}(k)$ are related by means of
\cite{Scatt-1d-1,Scatt-1d-2}
\begin{eqnarray}
T_{l=0}(k) = -\frac{2\hbar^2}{m} \frac{1}{\cot[\delta(k)]-\text{i}}.
\label{T-1D}
\end{eqnarray}
So, inserting Eq. (\ref{C-1D}) into Eq. (\ref{T-general}), and comparing with
(\ref{T-1D}) we get,
\begin{eqnarray}
V_{l=0}(k) = -\frac{2 \hbar^2}{m} k\tan[\delta(k)].
\label{V-1D}
\end{eqnarray}
Inserting Eq. (\ref{Expansion-1D}) into Eq. (\ref{V-1D}) and taking into account
the Taylor expansion at low momentum of $V_{l=0}(k)$ where
$V_{l=0}(k)=g^{(0)} + g^{(2)} k^2 + ...$, we can obtain the $1d$ intra-species
interacting coupling constants
\begin{eqnarray}
g^{(0)}= -\frac{2\hbar^2}{ma}, \quad\quad\quad
g^{(2)}=-\frac{\hbar^2}{m}r.
\label{g0_1d}
\end{eqnarray}
In a similar way the inter-species coupling constants are
\begin{eqnarray}
g^{(0)}_{12}=-\frac{2\hbar^2}{ma_{12}}, \quad\quad\quad
g^{(2)}_{12}=-\frac{\hbar^2}{m}r_{12}.
\label{g0_12_1d}
\end{eqnarray}
From the mean-field grand-potential in Eq. (\ref{grand-symmetric}) and using the
coupling constants in the zero-range approximation to the interactions in 
Eqs. (\ref{g0_1d}) and (\ref{g0_12_1d}), the energy per particle is given as
\begin{eqnarray}
\frac{E_0}{N}=\frac{1}{4}\Big(1+\frac{a}{a_{12}}\Big)g^{(0)}n.
\label{E0-1D}
\end{eqnarray}
From Eq. (\ref{grand-pot}) with $d = 1$, and performing dimensional
regularization, the grand-potential is given by 
\cite{1D-3D-mixtures,Regularization}
\begin{eqnarray}
\frac{\Omega_g}{L}= -\frac{2}{3\pi} \Big(\frac{m}{\hbar^2}\Big)^{1/2}
\frac{\mu^{3/2}}{(1+\delta_+\mu)^{1/2}}
- \frac{2}{3\pi} \Big(\frac{m}{\hbar^2}\Big)^{1/2}
\frac{\lambda^{3/2}\mu^{3/2}}{(1+\delta_-\mu)^{1/2}}.
\end{eqnarray}
So the $1d$ total energy per particle including the effects of the nonuniversal 
corrections to the interactions can be read as
\begin{eqnarray}
\frac{E_{FR}}{N} &=& \frac{1}{4}\big(g^{(0)}+g^{(0)}_{12}\big)n
\nonumber \\
&-&\frac{1}{3\sqrt{2}\pi}\Big(\frac{m}{\hbar^2}\Big)^{1/2}
\frac{[g^{(0)}+g^{(0)}_{12}]^{3/2}}
{\big[1 + \frac{2m}{\hbar^2}(g^{(2)}+g^{(2)}_{12})n\big]^{3/2}}n^{1/2}
\nonumber \\
&-&\frac{1}{3\sqrt{2}\pi}\Big(\frac{m}{\hbar^2}\Big)^{1/2}
\frac{[g^{(0)}-g^{(0)}_{12}]^{3/2}}
{\big[1 + \frac{2m}{\hbar^2}(g^{(2)}-g^{(2)}_{12})n\big]^{3/2}}n^{1/2}.
\label{Total-energy-FR-1d}
\end{eqnarray}

As is known the stable formation of a droplet requires a subtle equilibrium
between attraction and repulsion \cite{Petrov1,Petrov2}.
In Eq. (\ref{Total-energy-FR-1d}) we can see that such a balance is achieved
providing $g^{(0)} + g^{(0)}_{12}>0$ and $g^{(0)} > g^{(0)}_{12}$, and the 
liquidlike phase remains stable throughout.
The first condition indicates that the mean-field energy per particle is 
repulsive and stable.
The LHY-like corrections including the nonuniversal effects of the interactions
are negative and these provide the necessary attractive contribution.
Moreover due to the second condition, the LHY-like correction does not undergo
from the issue of imaginary contributions, unlike the  $3d$ model.
Therefore we can use the full expression for the energy per particle without
any approach that requires removal any of terms.
Taking into account the above two mentioned conditions for a mixture with
$g^{(0)}>0$ and $g^{(0)}_{12}>0$ the leading contribution is very repulsive and 
the attractive subleading contribution is not enough to allow the
accommodation of a large number of atoms. Therefore predicted energies per
particle are around two orders of magnitude smaller than those we will discuss 
in the following.
Another way to obtain a balanced repulsion-attraction effect which leads to the
emergence of a stable $1d$ DNUC or an usual droplet with higher energies per 
particle, consisting of using $|g^{(0)}|>|g^{(0)}_{12}|$ with $g^{(0)}>0 $, and
$g^{(0)}_{12}<0$, which corresponds to repulsive intra-species and attractive 
inter-species interactions, respectively.
We focus on this last case. We use $g=2\hbar^2/m|a|$, and 
$g_{12}=-2\hbar^2/m |a_{12}|$.
We take $E_B=\hbar^2/ m |a|^2$ as energy unit, we define
$\epsilon=g^{(0)}_{12}/g^{(0)}<0$, $\bar{n}=n|a|$, $\bar{r}=r/|a|$, 
$\bar{r}_{12}=r_{12}/|a|$, and $\bar{\mathcal{E}}=E/(L E_B/|a|)$.
Thus these new variables lead to the scaled energy per particle of an uniform 
droplet with nonuniversal corrections to the interactions
$\bar{\mathcal{E}}_{DNUC}/\bar{n}$ as
\begin{eqnarray}
\frac{\bar{\mathcal{E}}_{DNUC}}{\bar{n}} =\frac{1}{2}(1-|\epsilon|)\bar{n}
- \frac{2}{3\pi} \sqrt{\bar{n}}\Big[
\frac{(1-|\epsilon|)^{3/2}}
{[1-2(\bar{r}+\bar{r}_{12})\bar{n}]^{3/2}}
+ \frac{(1+|\epsilon|)^{3/2}}
{[1-2(\bar{r}-\bar{r}_{12})\bar{n}]^{3/2}}
\Big].
\label{FR-E-1d}
\end{eqnarray}
Within Petrov's approximation, $|\epsilon|=1$ only in the fluctuations, we have
\begin{eqnarray}
\frac{\bar{\mathcal{E}}_{DNUC+P}}{\bar{n}} =\frac{1}{2}(1-|\epsilon|)\bar{n}
- \frac{4\sqrt{2}}{3\pi} \frac{\sqrt{\bar{n}}}
{[1-2(\bar{r}-\bar{r}_{12})\bar{n}]^{3/2}}.
\label{FR-P-1d}
\end{eqnarray}

\begin{figure}[t]
\begin{center}
\includegraphics[height=7cm, width=8cm,clip]{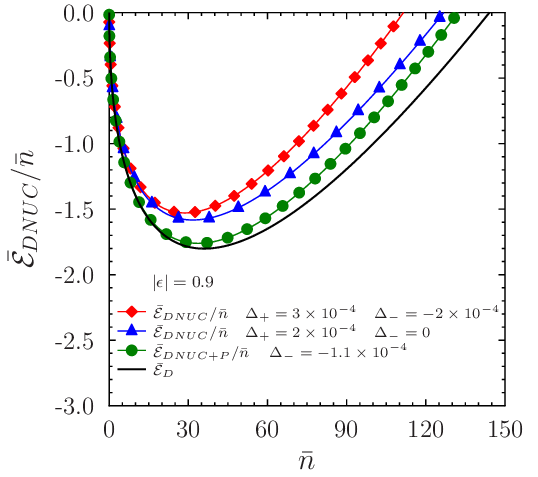}
\end{center}
\vskip -.8cm
\caption{One-dimensional energy per particle of droplets with nonuniversal 
corrections to the interactions as function of the density, Eq. (\ref{FR-E-1d}).
We set $|\epsilon| = 0.9$, and define $\Delta_\pm = \bar{r}\pm\bar{r}_{12}$.
We consider $\Delta_+ = 3\times 10^{-4}$ and $\Delta_- = -2\times 10^{-4}$
(red diamonds).
$\Delta_+ = 2\times 10^{-4}$ and $\Delta_- =  0$ (blue triangles).
In the simplified version of Eq. (\ref{FR-E-1d}) given by Eq. (\ref{FR-P-1d})
we consider $\Delta_+ = -1.1\times 10^{-4}$  (green circles).
We also include results in absence of nonuniversal contributions to the 
interactions in Eq. (\ref{FR-P-1d}) $\mathcal{E}_D$ (solid black line), given in 
Ref. \cite{Petrov2}.
}
\label{F6}
\end{figure}

In Fig \ref{F6}, we plot  Eq. (\ref{FR-E-1d}), and the simplified version given
by Eq. (\ref{FR-P-1d}). We consider $|\epsilon| = 0.9$, and different values of
the parameter $\Delta_\pm = \bar{r}\pm\bar{r}_{12}$.
We also include results in absence of nonuniversal corrections to the 
interactions from  Eq. (\ref{FR-P-1d}), $\bar{\mathcal{E}}_D$.
The values of $\Delta_{\pm}$ have been chosen to obtain lower energies per
particle compared to conventional droplets.
This behavior is in qualitatively agreement to Monte Carlo results  for
conventional droplets \cite{Petrov2}.
The repulsion-attraction balance that gives rise to $1d$ DNUC is metastable if 
the nonuniversal correction to the repulsion exceeds the nonuniversal correction
of the attraction.
Such  a balance is broken providing $\Delta_- \gtrsim 4.5 \times 10^{-4}$ in
Eq. (\ref{FR-P-1d}). The DNUC becomes unstable and it is not possible to achieve
a self-bound state.
In Eq. (\ref{FR-E-1d}) with $\Delta_->0$ there is a similar behavior, but the 
instability occurs faster due to the presence of the term with the factor
$\Delta_+$.

Using Eq. (\ref{Total-energy-FR-1d}) we obtain the improved $1d$ GPE describing
free droplets with nonuniversal corrections to the interactions as
\begin{eqnarray}
\text{i}\hbar \frac{\partial \phi}{\partial t}
&=& \Big[-\frac{\hbar^2}{2m} \frac{\partial^2}{\partial x^2}
+ \frac{1}{2} \delta g^{(0)}_-|\phi|^2 
- \frac{1}{2} \delta g^{(2)}_{+}\frac{\partial^2}{\partial x^2}|\phi|^2
\nonumber \\
&-& \frac{1}{2\sqrt{2}\pi}\Big(\frac{m}{\hbar^2}\Big)^{1/2} 
\frac{[\delta g^{(0)}_-]^{3/2}}{\big[1+\frac{2m}{\hbar^2} 
\delta g^{(2)}_{+}|\phi|^2\big]^{5/2}}
|\phi|
\nonumber \\
&-& \frac{1}{2\sqrt{2}\pi}\Big(\frac{m}{\hbar^2}\Big)^{1/2} 
\frac{[\delta g^{(0)}_+]^{3/2}}{\big[1+\frac{2m}{\hbar^2} 
\delta g^{(2)}_{-}|\phi|^2\big]^{5/2}}
|\phi|
\Big]\phi,
\end{eqnarray}
with $\delta g^{(0)}_{\pm} = |g^{(0)}| \pm |g^{(0)}_{12}| >0$, 
and $\delta g^{(2)}_{\pm} = g^{(2)} \pm g^{(2)}_{12}$.
The simplified version of this last expression taking into account the Petrov's
approximation is given by
\begin{eqnarray}
\text{i}\hbar \frac{\partial \phi}{\partial t}
&=& \Big[-\frac{\hbar^2}{2m} \frac{\partial^2}{\partial x^2}
+ \frac{1}{2} \delta g^{(0)}_-|\phi|^2 - \frac{1}{2} \delta g^{(2)}_{+}
\frac{\partial^2}{\partial x^2}|\phi|^2
\nonumber \\
&-& \frac{1}{\pi}\big(\frac{m}{\hbar^2}\big)^{1/2} [g^{(0)}]^{3/2}
\frac{|\phi|}{\big[1+\frac{2m}{\hbar^2} \delta g^{(2)}_{-}|\phi|^2\big]^{5/2}}
\Big]\phi,
\end{eqnarray}
From Eqs. (\ref{g0_1d}) and (\ref{g0_12_1d}) we can see that at level of 
mean-field approximation the nonuniversal contribution to the interactions 
provides a repulsive contribution.
Increasing the density could be possible to find a critical value of
$\delta g^{(2)}_\pm$ such that the mixture becomes unstable.
This instability arises providing that the density of particles satisfies
$|\phi|^2 = 1/[2(r\pm r_{12})]$, with $r>r_{12}$ choosing the negative sign.
Therefore these two improved GPEs could be also useful to understanding the role
of nonuniversal corrections to the interactions into properties of dark and 
bright solitons.

\subsection{One-dimensional finite-temperature effects}

In the continuum limit the $1d$ one-loop correction to the grand potential in
expression (\ref{Grand-T}) leads to  \cite{1D-3D-mixtures}
\begin{eqnarray}
\frac{\Omega_{g}^{(T)}}{L}
= -\frac{1}{\pi} \sum_{\pm}\int_0^{\infty} dE_{\pm}(k_{\pm})
\frac{k_{\pm}}{e^{\beta E_{\pm}(k)}-1},
\end{eqnarray}
with $k^2_{\pm}$ given by  Eqs.~(\ref{k+}) and ~(\ref{k-}).
After introducing the variables $x_{\pm}= \beta E_{\pm}$, and expanding
$k_{\pm}$ at low temperature $k_B T \ll |\mu|$ \cite{Pit-Strin,Peth-Smith},
the $1d$ grand-canonical potential can be read as

\begin{eqnarray}
\frac{\Omega^{(T)}_g}{L}
&=& - \frac{\pi}{6}\Big(\frac{m}{\hbar^2}\Big)^{1/2}
\frac{(k_B T)^2}{\mu ^{1/2}} \Big[ 1- \frac{\pi^2}{20}
\frac{(1+\alpha_+ \mu)}{\mu^2}(k_B T)^2\Big]
\nonumber \\
&-&\frac{\pi}{6}\Big(\frac{m}{\hbar^2}\Big)^{1/2}
\frac{(k_B T)^2}{\lambda^{1/2}\mu ^{1/2}} \Big[ 1- \frac{\pi^2}{20}
\frac{(1+\alpha_- \mu)}{\lambda^2\mu^2}(k_B T)^2\Big],
\end{eqnarray}
and the respective thermal contribution to the energy per particle is
\begin{eqnarray}
\frac{E^{(T)}}{N} &=& \frac{\sqrt{2}\pi}{12} \Big( \frac{m}{\hbar^2} \Big)^{1/2}
\frac{(k_B T)^2}{[g^{(0)} +  g^{(0)}_{12}]^{1/2}n^{3/2}} (1+h^+_{NUC})
\nonumber \\
&+&\frac{\sqrt{2}\pi}{12} \Big( \frac{m}{\hbar^2} \Big)^{1/2} 
\frac{(k_B T)^2}{[g^{(0)} -  g^{(0)}_{12}]^{1/2}n^{3/2}} (1+h^-_{NUC}),
\label{Energy-T-1d}
\end{eqnarray}
with
\begin{eqnarray}
h^{\pm}_{NUC} = -\frac{\pi^2}{5}
\frac{(k_B T)^2}{[g^{(0)} \pm  g^{(0)}_{12}]^{2}n^{2}}
\Big[1+ \frac{6m}{\hbar^2}(g^{(2)} \pm  g^{(2)}_{12})n \Big].
\end{eqnarray}
Clearly the $1d$ energy per particle in Eq. (\ref{Energy-T-1d}) is divergent
providing $g^{(0)} = \pm g^{(0)}_{12} $. For attractive inter-species
interactions lightly larger than repulsive intra-species interaction this energy 
per particle becomes complex, just like in $2d$ and $3d$ mixtures.
In this sense we have a thermal-induced instability. 
However, even more remarkably instead both $1d$ DNUC and usual $1d$ droplets 
need to satisfy $g^{(0)} + g^{(0)}_{12}>0$ and holds 
$g^{(0)} > \pm g^{(0)}_{12}$. Therefore both are stable at finite temperature.
Currently, the region of greatest interest in the study of droplets is around
$-g_{12} \sim g $, so expression (\ref{Energy-T-1d}) is simplified dropping out
to the second line.
We also mention that the nonuniversal corrections to the interactions appear only
up at order $T^4$ and, hence at very low temperatures these could be neglected.

\section{Summary and outlook}
\label{Summary and outlook}

In the framework of a quantum effective field theory we derive the equation of
state at zero and low temperature of $3d$, $2d$ and $1d$ ultracold Bose-Bose 
mixtures of alkali-metal atoms considering nonuniversal corrections to the
interactions.
We perform one-loop Gaussian functional integration.
The zero-point energy ultraviolet divergence is removed through dimensional
regularization in $3d$ and in $1d$.
In $2d$ we deal this issue by means of dimensional regularization in the 
modified minimal subtraction scheme.
We stress and focus our analysis on the study of some regimes which the mixture
manifests the existence and stable formation of a liquidlike phase with 
nonuniversal corrections to the interactions.
In particular, when the nonuniversal contribution to the attraction exceeds that 
of repulsion in $3d$ DNUC, it results in an increasing in the equilibrium density 
regarding to the theoretical model to describe conventional droplets.
The increase in the nonuniversal effect to the attraction causes the droplet to 
be unstable, and it disappears.
The opposite happens when the nonuniversal correction to the attraction is 
exceeded by the repulsive counterpart.
The equilibrium density and the respective minimum energy are lower than 
conventional droplet and there is not instability.
If we consider the nonuniversal contributions as fitting parameters we found a 
good agreement from our model with some DMC results.
In $2d$ DNUC we have found small deviations from the equilibrium density
regarding usual droplets.
A possible explanation for this could be associated with the coupling constants
used. Due the perturbative nature of the self-bound liquidlike phase could be
suitable consider coupling constants for $a\ll r$.
In the $1d$ DNUC we obtain a qualitatively agreement with the trend of Monte
Carlo results for conventional droplets.
A clear extension of this work is verify the validity of our results by means of
quantum Monte Carlo techniques
\cite{Montecarlo-Bose-Bose1,Montecarlo-Bose-Bose2,Montecarlo-Bose-Bose3}.
We also propose some improved GPEs in $3d$, $2d$ and $1d$ including the enhanced
nonuniversal contribution to the interactions.
In these equations the parameter handling could be suitable for fitting with
experimental or numerical evidence.
Likewise, we consider that these equations could be useful in the study of the 
effects of the nonuniversal corrections to the interactions in paradigms that 
apply broadly to many-body quantum systems, as solitons and vortices. 
Given the experimental relevance of harmonic trapping potentials to achieve
quasi-$2d$ or quasi-$1d$ regimes, it would be interesting not only to study
harmonically trapped Bose-Bose mixtures including the nonuniversal
corrections to the interactions, but also properties of DNUC at dimensional 
crossovers.
We also consider that our theoretical predictions could stimulate experimental
breakthroughs in the study in its own right of Gaussian fluctuations with 
nonuniversal corrections to the interactions in Bose-Bose mixtures.
Sizable nonuniversal effects could be reached experimentally through 
Feshbach-resonance technique by increasing the density and decreasing the 
scattering length.

Finite-temperature effects deserve a further study on their own right.
In principle in the droplet phase with or without nonuniversal corrections to the
interactions both in $3d$ and in $2d$, one could have the presence of a phase
transition, even if the mixture is stable at zero temperature.
For the $1d$ mixture could be interesting to see the thermal mechanisms driving 
the liquid-to-gas transition introduced in \cite{Evaporation}.
As our finite-temperature results are general for a symmetric Bose-Bose mixture 
could be interesting see how behaves the miscibility-separation condition 
regarding zero-temperature results \cite{Pit-Strin,Peth-Smith}.
Our finite-temperature results could be contrasted with the recently implemented
Monte Carlo methods at finite temperature
\cite{Montecarlo-finite-T-1,Montecarlo-finite-T-2,Montecarlo-finite-T-3}.

In $3d$ we consider two different expansions of the phase shift, the conventional
one, given in Eq. (\ref{fr}) and holding $a\gg r$ and a less known in 
Eq. (\ref{fr-2}) for $a\ll r$. 
From a scattering perspective it would be interesting to have expressions in the 
regime $a\ll r$ for both $2d$ and $1d$, and also consider their effects on the 
liquidlike phase with nonuniversal corrections to the interactions.

\section*{Acknowledgments}

I acknowledge partial support by Coordena\c{c}\~ao de Aperfei\c{c}oamento de
Pessoal de N\'ivel Superior CAPES (Brazil).
I thank Professor Sadhan Adhikari for recommendation on using of expansion
(\ref{fr-2}), and comments on a preliminary version of the present work.

\end{document}